\documentclass{zhuang}

\begin{document}

\markboth{} { Phase Structure of Color Superconductivity }

%
\catchline{}{}{}{}{}
%

\title{Phase Structure of Color Superconductivity and Chiral Restoration\\
}

\author{ Pengfei Zhuang}

\address{Physics Department, Tsinghua University\\ Beijing 100084, China
\\
zhuangpf@mail.tsinghua.edu.cn}

\maketitle

\begin{abstract}
We investigate color superconductivity and chiral symmetry
restoration at finite temperature and baryon density in the frame
of standard two flavor Nambu--Jona-Lasinio model. We derive the
diquark mass in RPA, discuss its constraint on the coupling
constant in the diquark channel, and find a strong competition
between the two phase transitions when the coupling constant is
large enough.

\end{abstract}

\ccode{PACS Nos.: 11.30.Rd, 12.38.Lg, 11.10.Wx, 25.75.Nq}

\section{Introduction}
In the ideal case at asymptotically high baryon density, the color
superconductivity with two massless flavors and the
color-flavor-locking (CFL) phase with three degenerated massless
quarks have been widely discussed from first principle QCD
calculations\cite{cfl}. For physical applications we are more
interested in the moderate baryon density region which may be
related to the neutron stars and, in very optimistic cases, even
to heavy-ion collisions. Usually, four-fermion interaction models
such as the instanton, as well as the Nambu--Jona-Lasinio (NJL)
model, are used in the study of color superconductivity phase
transition. It was found that the color superconductivity gap can
be of the order of 100 MeV\cite{csc1}, which is two orders larger
than early perturbative QCD estimation\cite{csc2}. The electric
and color charge neutrality condition\cite{bp2-1,bp2-2} leads to a
new phase of color superconductivity, the gapless color
superconductivity\cite{gapless} or the breached pairing
phase\cite{bp3}. The most probable temperature for this new phase
is finite but not zero\cite{bp4}.

It is generally accepted that the NJL model\cite{njl1} applied to
quarks\cite{njl2} offers a simple but effective scheme to study
spontaneous chiral symmetry breaking in the vacuum, chiral
restoration at finite temperature and density, and spontaneous
color symmetry breaking at high
density\cite{bp2-2,gapless,bp4,njl3}. From the study of chiral
phase transition without considering color superconductivity, it
is well-known that\cite{njl2} the mean field approximation to
quarks and random phase approximation (RPA) to mesons can describe
well the thermodynamics of the system, especially the massless
Goldstone mode in the spontaneous symmetry breaking phase. In this
letter I will determine the diquark mass in RPA, discuss its
constraint on the coupling constant in the diquark channel, and
investigate the phase structure of color superconductivity and
chiral restoration and its dependence on the coupling constant.

\section{Quarks in Mean Field Approximation}
The NJL Lagrange density is defined as
\begin{eqnarray}
\label{njl} {\cal L} &=&
\bar{\psi}\left(i\gamma^{\mu}\partial_{\mu}-m+\mu\gamma_0\right)\psi\\
&+&G_S\left(\left(\bar{\psi}\psi\right)^2+\left(\bar{\psi}i{\bf
\tau}\gamma_5\psi\right)^2
\right)+G_D\left(\bar\psi^C_{i\alpha}\epsilon^{ij}\epsilon^{\alpha\beta
\gamma}i\gamma^5\psi_{j\beta}\right)\left(\bar\psi_{i\alpha}\epsilon^{ij}\epsilon^{\alpha\beta
\gamma}i\gamma^5\psi^C_{j\beta}\right) \ ,\nonumber
\end{eqnarray}
where $G_S$ and $G_D$ are coupling constants in color signet
channel and color anti-triplet channel, respectively, and $m$ is
the current quark mass.

The quark-antiquark and diquark condensates which are order
parameters of chiral and color superconductivity phase transitions
are defined as
\begin{equation}
\label{condensate} \sigma = \langle\bar\psi\psi\rangle\ ,\ \ \
\Delta = \Delta^3 = \langle
\bar\psi^C_{i\alpha}\epsilon^{ij}\epsilon^{\alpha\beta
3}i\gamma^5\psi_{j\beta}\rangle =  \langle
\bar\psi_{i\alpha}\epsilon^{ij}\epsilon^{\alpha\beta
3}i\gamma^5\psi^C_{j\beta}\rangle\ ,
\end{equation}
where it has been regarded that only the first two colors
participate in the condensate, while the third one does not. The
condensates are assumed real but are otherwise as yet unspecified.

A usual way to treat systematically the spontaneous color and
chiral symmetry breaking at finite temperature and density is to
introduce the Nambu-Gorkov space\cite{ng} with color and flavor
degrees of freedom. After performing the Fourier transformation,
the quark propagator in momentum space at mean field level is
diagonal in the $12$ dimensional Nambu-Gorkov space,
\begin{equation}
\label{propagator} S(k)= \left(\begin{array}{cccccc} S_A(k)& & & &
&
\\ & S_B(k) & & & & \\ & & S_C(k) & & & \\ & & &
S_D(k) & & \\ & & & & S_E(k) & \\ & & & & &
S_F(k)\end{array}\right)\ ,
\end{equation}
where $S_I(k)$ with $I=A, B, C, D, E, F$ are $2\times 2$ matrices,
\begin{equation}
\label{matrix}
S_I = \left(\begin{array}{cc} G_I^+& \Xi_I^-\\
\Xi_I^+ & G_I^- \end{array}\right) \ .
\end{equation}
In general case, the quark chemical potential $\mu_{i\alpha}$
depends on color and flavor degrees of freedom, the elements of
the $6$ matrices $S_I$ are totally different. However, if we
consider only the baryon chemical potential $\mu_B$, one has $\mu
= \mu_B/3$, and some of the elements are identical ,
\begin{eqnarray} \label{element}
&& G_A^+ = G_B^+ = G_C^+ = G_D^+\ ,\ \ \ G_A^- = G_B^- = G_C^- = G_D^-\ ,\\
&& \Xi_A^+ = \Xi_B^+ = -\Xi_C^+ = -\Xi_D^+\ ,\ \ \ \Xi_A^- = \Xi_B^- = -\Xi_C^- = -\Xi_D^-\ ,\nonumber\\
&& G_E^+ = G_F^+\ ,\ \ \ G_E^- = G_F^-\ ,\ \ \ \Xi_E^+ = \Xi_F^+ = \Xi_E^- = \Xi_F^- = 0\ ,\nonumber\\
&& G_A^+(k) = {k_0+E_k-\mu\over
(k_0-E_k^-)(k_0+E_k^-)}\Lambda_+\gamma_0+
{k_0-E_k-\mu\over (k_0-E_k^+)(k_0+E_k^+)}\Lambda_-\gamma_0\ ,\nonumber\\
&& G_A^-(k) = {k_0-E_k+\mu\over
(k_0-E_k^-)(k_0+E_k^-)}\Lambda_-\gamma_0+
{k_0+E_k+\mu\over (k_0-E_k^+)(k_0+E_k^+)}\Lambda_+\gamma_0\ ,\nonumber\\
&& \Xi_A^+(k) = {2iG_D\Delta\over
(k_0-E_k^-)(k_0+E_k^-)}\Lambda_-\gamma_5+
{2iG_D\Delta\over (k_0-E_k^+)(k_0+E_k^+)}\Lambda_+\gamma_5\ ,\nonumber\\
&&\Xi_A^-(k) = {2iG_D\Delta\over
(k_0-E_k^-)(k_0+E_k^-)}\Lambda_+\gamma_5+ {2iG_D\Delta\over
(k_0-E_k^+)(k_0+E_k^+)}\Lambda_-\gamma_5\ ,\nonumber\\
&& G_E^+(k) = {1\over k_0-E_k +\mu}\Lambda_+\gamma_0+
{1\over k_0+E_k+\mu}\Lambda_-\gamma_0\ ,\nonumber\\
&& G_E^-(k) = {1\over k_0-E_k-\mu}\Lambda_+\gamma_0+ {1\over
k_0+E_k-\mu}\Lambda_-\gamma_0\ ,\nonumber
\end{eqnarray}
with the quark energies $E_k^\pm = \sqrt{\left(E_k\pm
\mu\right)^2+\left(2G_D\Delta\right)^2}\ ,\ E_k = \sqrt{|{\bf
k}|^2+M_Q^2}$, effective quark mass $M_Q = m-2G_S\sigma\ ,$ and
the energy projectors $\Lambda_{\pm} = {1\over
2}\left(1\pm{\gamma_0\left({\bf \gamma\cdot k}+M_Q\right)\over
E_k}\right)\ .$

While the first two colors are involved in the quark-antiquark and
diquark condensates, and their dispersion relation $E_k^\pm$ are
related to $\sigma$ and $\Delta$, the third color participates in
the chiral condensate only, the non-diagonal elements of the last
two matrices $S_E$ and $S_F$ are zero, and its energy $E_k$ is
associated with $\sigma$ only.

From the definition of the two condensates (\ref{condensate}), it
is easy to obtain their relation with the elements of the quark
propagator (\ref{propagator}),
\begin{eqnarray}
\label{gap} && \sigma=4M_Q\int{d^3\bf k\over (2\pi)^3}{1\over
E_k}\Big[{E_k-\mu\over E_k^-}\left(2f(E_k^-)-1\right)\\
&&\ \ \ +{E_k+\mu\over E_k^+}\left(2f(E_k^+)
-1\right)+f(E_k-\mu)+f(E_k+\mu)-1\Big]\ ,\nonumber\\
&& \Delta\left[1+8G_D\int{d^3{\bf k}\over (2\pi)^3}\left({1\over
E_k^-}\left(2f(E_k^-)-1\right)+{1\over
E_k^+}\left(2f(E_k^+)-1\right)\right)\right] = 0\ ,\nonumber
\end{eqnarray}
with the Fermi-Dirac distribution function $ f(x) = 1/(
e^{x/T}+1)$.

\section{ Coupling Constant $G_D$ in Diquark Channel}
The diquark polarization function can be represented in terms of
the matrix elements of the quark propagator,
\begin{equation}
\label{dpolari} -i\Pi_D(k) = -4\int{d^4p\over
(2\pi)^4}Tr\left[i\gamma_5 iG_A^+(p-k)i\gamma_5 iG_A^-(p)\right]\
.
\end{equation}
In the mean field approximation to quarks and RPA to mesons and
diquarks, the temperature and baryon chemical potential dependence
of the diquark mass $M_D(T,\mu)$ is controlled by the pole
equation
\begin{equation}
\label{dmass} 1-2G_D\Pi_D(k_0^2=M_D^2,{\bf 0}) = 0\ .
\end{equation}

We consider now the physical constraints in the vacuum on the
coupling constant $G_D$ in the diquark channel. The diquark mass
$M_D(0,0)$ in the vacuum is determined by
\begin{equation}
\label{dvacuum} 1-8G_D\int {d^3{\bf p}\over (2\pi)^3}\left({1\over
E_p +M_D(0,0)/2}+{1\over E_p -M_D(0,0)/2}\right)=0\ .
\end{equation}

In general, temperature effect disorders a system, and any
condensate will be suppressed in hot medium. Therefore, the
critical chemical potential of color superconductivity goes up
monotonously with increasing temperature, and the minimum one
$\mu_\Delta$ at zero temperature satisfies the gap equation
\begin{equation}
\label{critical} 1-8G_D\int {d^3{\bf p}\over
(2\pi)^3}\left({1\over E_p +\mu_\Delta}+{1\over E_p
-\mu_\Delta}\right)=0\ .
\end{equation}
From the comparison of the above two equations, we obtain the
relation between the diquark mass in the vacuum and the critical
chemical potential for diquark condensate at zero temperature,
$\mu_\Delta = M_D(0,0)/2\ .$

When the diquark is massless in the vacuum, the color
superconductivity happens even at $\mu_\Delta =0$, and the vacuum
becomes instable. On the other hand, when the diquark mass in the
vacuum is larger than two times the quark mass, the color
superconductivity is far from the vacuum, and a diquark can not be
in the bound state, but decay into two quarks. Therefore, the
diquark mass $M_D(0,0)$ should satisfy the constraint $0 <
M_D(0,0) < 2M_Q(0,0)\ .$ From the comparison of the diquark mass
equation (\ref{dvacuum}) and the quark mass equation in the
vacuum,
\begin{equation}
\label{mass}
1-24G_S\int{d^3{\bf p}\over (2\pi)^3} {1\over
\sqrt{M_Q^2(0,0)+{\bf p}^2}} = 0\ ,
\end{equation}
the constraint $0 < M_D(0,0) < 2M_Q(0,0)$ on the diquark mass
resulus in the relation between the two coupling constants $G_D$
and $G_S$,
\begin{eqnarray}
\label{c2}
&& G_D^{min} < G_D < G_D^{max}\ ,\nonumber\\
&& G_D^{min} = {\pi^2\over
4\left(\Lambda\sqrt{M_Q^2(0,0)+\Lambda^2}+M_Q^2(0,0)\ln{\Lambda+\sqrt{M_Q^2(0,0)+\Lambda^2}
\over M_Q(0,0)}\right)}\ ,\nonumber\\
&& G_D^{max} = {3\over 2}G_S {M_Q(0,0)\over M_Q(0,0)-m}\ .
\end{eqnarray}
With the known values of the parameters $\Lambda, m$ and
$M_Q(0,0)$ in real world, the minimum and maximum coupling
constant take the values $G_D^{min} = 0.82 G_S$ and
$G_D^{max}=1.55 G_S$.

\section{Phase Diagram }
Before making numerical calculations, we first discuss the
parameters in the model. There are four independent parameters in
the current NJL model with both diquark and meson channels, the
three-momentum cutoff $\Lambda$, the current quark mass $m$, and
the coupling constants $G_S$ and $G_D$. The first three can be
fixed by fitting the pion mass $m_\pi =0.134$ GeV, the pion decay
constant $f_\pi = 0.093$ GeV, and the quark condensate density
$\sigma = (-0.25$ GeV$)^3$ in the vacuum. It leads to $\Lambda =
0.65$ GeV and $G_S=5.01$ GeV$^{-2}$ in chiral limit with zero
current quark mass $m=0$, and $\Lambda = 0.653$ GeV, $G_S=4.93$
GeV$^{-2}$ and $m=0.005$ GeV in real world.

\begin{figure}[ht]
\vspace*{+0cm} \centerline{\epsfxsize=2.9in
\epsffile{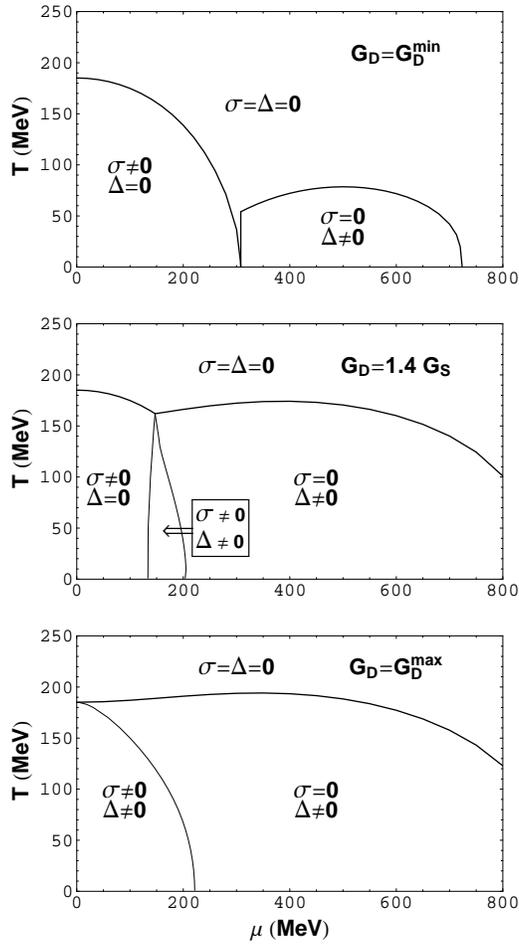}} \caption{\it The Phase diagram in the
$T-\mu$ plane for different coupling constant $G_D$ in the diquark
channel.} \label{fig1}
\end{figure}

In chiral limit the chiral phase transition is well defined, and
the phase transition line determined by $\sigma =0$ is a solution
of the gap equations (\ref{gap}). The phase diagram in the $T-\mu$
plane is shown in Fig.\ref{fig1} for different coupling constant
$G_D$ in the diquark channel.

At the minimum coupling constant $G_D^{min}$, the whole $T-\mu$
plane is separated into three regions, the region with chiral
symmetry breaking but color symmetry, $\sigma\ne 0$ and $\Delta
=0$ at low density, the region with chiral symmetry but color
symmetry breaking, $\sigma =0$ and $\Delta \ne 0$ at low
temperature, and the region with both chiral and color symmetries,
$\sigma =\Delta =0$ at high temperature and/or high density. The
critical temperature $T_\chi = 185$ MeV for chiral phase
transition at $\mu =0$ is in good agreement with the lattice
calculation\cite{karsch,LA2}, and the chiral symmetry breaking
phase ends at $\mu_\chi$ determined by
\begin{equation}
\label{muc} 1-24G_S\int{d^3{\bf p}\over (2\pi)^3}{1\over
p}\theta(k-\mu_\chi) = 0\ .
\end{equation}
From the comparison with the quark mass equation (\ref{mass}) in
the vacuum, one has $\mu_\chi = M_Q(0,0)$. Therefore, from the
definition of the minimum coupling constant $G_D^{min}$ at which
the minimum chemical potential $\mu_\Delta$ of color
superconductivity is equal to the constituent quark mass in the
vacuum, the end point of chiral phase transition and the starting
point of color superconductivity coincide at $\mu = M_Q(0,0)$.

The order of a phase transition is determined by the behavior of
the order parameter across the phase transition line. It is well
known\cite{njl2} that the chiral phase transition is of second
order at high temperature and of first order at high density. The
critical chemical potential at which the chiral transition goes
from first to second order is $285$ MeV in the NJL
model\cite{njl6}. One can also determine the critical exponent for
the color superconductivity. Since the square bracket of the
second gap equation of (\ref{gap}) depends quadratically on the
order parameter $\Delta$, one has a simple zero for $\Delta^2$ as
a function of $T$ and therefore in the neighborhood of the
critical temperature $T_\Delta$ of color superconductivity the
order parameter behaviors as $\Delta(T) \sim |T-T_\Delta|^{1/2}\
.$ This means that the transition between the phase with
$\sigma=0, \Delta\ne 0$ and the phase with $\sigma=\Delta=0$ is of
second order.

With increasing coupling constant $G_D$, the critical temperature
$T_\Delta$ for melting the diquark condensate increases and the
critical chemical potential $\mu_\Delta$ for forming the color
superconductivity decreases. Especially, a new phase with both
chiral and color symmetry breaking, $\sigma\ne 0$ and
$\Delta\ne0$, is produced in between the phase with only chiral
symmetry breaking and the phase with only color symmetry breaking,
as shown in Fig.\ref{fig2} for $G_D = 1.4 G_S$. The left and right
borders of this mixed phase are determined by $\sigma\ne 0, \Delta
=0$ and $\sigma =0, \Delta \ne 0$, respectively. With increasing
$G_D$ the chiral breaking phase is swallowed up gradually by the
mixed phase. At the maximum coupling constant $G_D^{max}$ the
chiral breaking phase disappears and the mixed phase reaches the
maximum.

\section{Summary}
We have analytically derived the simple relation between the
diquark mass in the vacuum and the minimum quark chemical
potential for diquark condensate, $\mu_\Delta = M_D(0,0)/2$. Under
the conditions to keep the diquark bound state in the vacuum and
to have a stable physical vacuum without diquark condensate in it,
we obtained the constraint on the coupling constant in the diquark
channel, $G_D^{min} < G_D < G_D^{max}$, with the relations to the
coupling constant in the meson channel $G_D^{min} \simeq 0.82 G_S$
and $G_D^{max}\simeq 3/2G_S$. There is a strong competition
between the color superconductivity and chiral symmetry
restoration: With increasing $G_D$ the phase with chiral symmetry
breaking but color symmetry is gradually taken over by the phase
with both chiral and color symmetry breaking.

\section*{Acknowledgments}
I thank Lianyi He and Meng Jin for the help in numerical
calculations. The work was supported in part by the grants NSFC
10425810, 10435080, and G2000077407.


\begin{thebibliography}{0}

\bibitem{cfl} M.Alford, K.Rajapopal, and F. Wilczek, Nucl. Phys. {bf B537}, 443(1999);
T.Sch\"afer and F.Wilczek, Phys. REv. Lett. {bf 82},
3956(1999);T.Sch\"afer, Nucl. Phys. {\bf B575}, 269(2000);
D.Rischke, Phys. Rev. {\bf D62}, 054017(2000).

\bibitem{csc1} R.Rapp, T.Sch\"afer, E..V.Shuryak, and M.Velkovsky,
Phys. Rev. Lett. {81}, 53(1998); M.Alford, K.Rajagopal, and
F.Wilczek, Phys. Lett. {\bf B422}, 247(1998).

\bibitem{csc2} D.Bailin and A.Love, Phys. Rep.{\bf 107}, 32591984).

\bibitem{bp2-1} M.Alford and K.Rajagopal, JHEP06:031(2002);
A.W.Steiner, S.Reddy, and M.Prakash, Phys. Rev. {\bf D66},
094007(2002).

\bibitem{bp2-2} M.Huang, P.Zhuang, and W.Chao, Phys. Rev. {\bf D67}, 065015(2003).

\bibitem{gapless} I. Shovkovy and M. Huang, Phys. Lett. {\bf B564} 205(2003).

\bibitem{bp3} K.Rajagopal and F.Wilczek, Phys. Rev. Lett. {\bf 86} 3492(2001);
F.Neumann, M.Buballa, and M.Oertel, Nucl. phys. {\bf
A714}481(2003); W.V.Liu and F.Wilczek, Phys. Rev. Lett.{\bf 90},
047002(2003).

\bibitem{bp4} J.Liao and P.Zhuang, Phys. Rev. {\bf D68},
114016(2003).

\bibitem{njl1} Y.Nambu and G.Jona-lasinio, Phys. Rev. {\bf
122}, 345(1961); {\bf 124}, 246(1961).

\bibitem{njl2} See, for reviews and general references, U.Vogl and W.Weise, Prog.
Part. and Nucl. Phys. {\bf 27}, 195(1991); S.P.Klevansky, Rev.
Mod. Phys. {\bf 64}, 649(1992); M.K.Volkov, Phys. Part. Nucl. {\bf
24}, 35(1993); T.Hatsuda and T.Kunihiro, Phys. Rep. {\bf 247},
338(1994).

\bibitem{njl3}  T.M.Schwarz, S.P.Klevansky, and G.Rapp, Phys. Rev. {\bf C60}, 055205(1999);
M.Huang, P.Zhuang, and W.Chao, Phys.Rev.{\bf D65}, (2002)076012.

\bibitem{ng} A.A.Abrikosov, L.P.Gorkov, and I.E.Dzyaloshinski,
Method of Quantum Field Theory in Statistical Physics, 1963;
A.L.Fetter and J.D.Walecka, Quantum Theory of Many-Particle
Systems, 1971; D.Rischke, Phys. Rev. {\bf D62}, 034007(2000).

\bibitem{karsch} See, for example, F.Karsch, Lect. Notes Phys. {\bf 583},209(2002).

\bibitem{LA2}     {X.Q.Luo, Phys.Rev.{\bf D70},(2004)091504(R);
H.S.Chen and X.Q.Luo, hep-lat/0411023; E.B.Gregory, S.H.Guo, H.Kroger and X.Q.Luo,
Phys.Rev.{\bf D62}, (2000)054508.}

\bibitem{njl6} J.H\"ufner, S.P.Klevansky, P.Zhuang and H.Voss,
Ann. Phys. (N.Y.) {\bf 234}, 225(1994); P.Zhuang, J.H\"ufner and
S.P.Klevansky, Nucl. Phys. {\bf A576}, 525(1994).

\end{thebibliography}
\end{document}